\documentclass[aps,prb,twocolumn,superscriptaddress,showpacs,amsmath,amssymb,amsfonts,floatfix]{revtex4}

\usepackage{graphicx}
\usepackage{bm}

\begin{document}

\title{Chiral magnetic effect in the absence of Weyl node}

\author{Ming-Che Chang}
\affiliation{Department of Physics, National Taiwan Normal University, Taipei 11677, Taiwan}

\author{Min-Fong Yang}
\email{mfyang@thu.edu.tw}
\affiliation{Department of Applied Physics, Tunghai University, Taichung 40704, Taiwan}

\date{\today}

\begin{abstract}
The nodal points in a Weyl semimetal are generally considered as the causes of the chiral anomaly and the chiral magnetic effect (CME). Employing a linear-response analysis of a two-band lattice model, we show that the Weyl nodes and thus the chirality are not required for the CME, while they remain crucial for the chiral anomaly.
Similar to the anomalous Hall effect, the CME results directly from the Berry curvature of energy bands, even when there is no monopole source from the Weyl nodes. Therefore, the phenomenon of the CME could be observed in a wider class of materials.
Motivated by this result, we suggest that the nodeless CME may appear in three-dimensional quantum anomalous Hall insulators, but after they become metallic due to  the band deformation caused by inversion symmetry breaking.
\end{abstract}

\pacs{
71.90.+q, 
75.47.$-$m,  
03.65.Vf, 
73.43.$-$f 
}

\maketitle

\section{Introduction}
Materials with topologically-nontrivial electronic structure have recently been under intensive investigation. A particularly interesting state of matter is the three-dimensional Weyl semimetal,~\cite{Burkov2011a,Wan2011,Yang2011,Turner_Vishwanath2013} whose band structure contains isolated band-touching points, called Weyl nodes.
Such nodal points behave as sources/drains of the Berry flux and carry nonzero monopole charges $Q$, which is defined by the integral of the Berry curvature $\mathbf{\Omega}_\mathbf{k}$ over the surface enclosing the node, $Q=(1/2\pi)\oint\!\mathrm{d}\mathbf{S}_\mathbf{k}\cdot\mathbf{\Omega}_\mathbf{k}$.
The Weyl nodes are protected topologically against perturbations. They would disappear only if two nodes with opposite monopole charges merge and annihilate with each other. Recently, TaAs and NbAs have been experimentally confirmed to be Weyl semimetals.~\cite{Xu_a2015,Lv_a2015,Lv_b2015,Yang_etal2015,Xu_b2015} This progress paves the way for exploring novel effects in these materials.

Due to the nontrivial topology in momentum space, such nodal materials exhibit a wide variety of unusual electromagnetic responses.~\cite{rev_transport} In a Weyl semimetal, electrons near each Weyl node can be assigned a chirality by the monopole charge of that node. Applying a pair of non-orthogonal electric and magnetic fields, the charges can be transported between two Weyl nodes with opposite chiralities. Therefore, the number of electrons with a definite chirality is no longer conserved, showing the so-called chiral anomaly.~\cite{ABJ_anomaly} By using a semiclassical analysis, one can show that the anomalous source term in the continuity equation of chiral charges is proportional to the monopole charge.~\cite{Son_Yamamoto2012} Thus this exotic phenomenon  becomes more manifest for nodes with larger $Q$'s. This anomaly is predicted to give an enhanced negative magnetoresistance when the applied electric and magnetic fields are parallel to each other.~\cite{Nielsen_Ninomiya,Son_Spivak2013} Such a prediction has just been confirmed by experiments in the Weyl semimetal TaAs.~\cite{Huangx2015,Zhang2015}

Besides chiral anomaly, such nodal materials may show chiral magnetic effect (CME) when the energies of pairs of Weyl nodes are different.~\cite{Zyuzin_Burkov2012,Goswami_Tewari2013,Stephanov_Yin2012,
Son_Yamamoto2012,Zyuzin_etal2012} This effect gives a dissipationless electric current $\mathbf{J}$ flowing along an applied magnetic field $\mathbf{B}$, ${\bf J}=-\alpha {\bf B}$. Employing a low-energy effective theory with unbounded linear dispersion, the CME coefficient $\alpha$ is shown to be proportional to the energy separation between a pair of Weyl nodes. Since the CME can be related with the chiral anomaly through the energy balance of chirality generation,~\cite{Nielsen_Ninomiya,Fukushima_etal2008} one may expect that, when there is no Weyl node such that chiral fermions become ill-defined, neither the chiral anomaly nor the CME could exist.

However, most of the early investigations are based on effective models with linear dispersions around Weyl nodes. In a previous work, we find that, when going beyond the linear regime such that the concept of chirality may no longer be appropriate, the CME can still exist.~\cite{Chang_Yang2015} Our analysis shows that this effect is better understood in terms of the Berry curvature, rather than the chirality.

\begin{figure}
\includegraphics[width=3.3in]{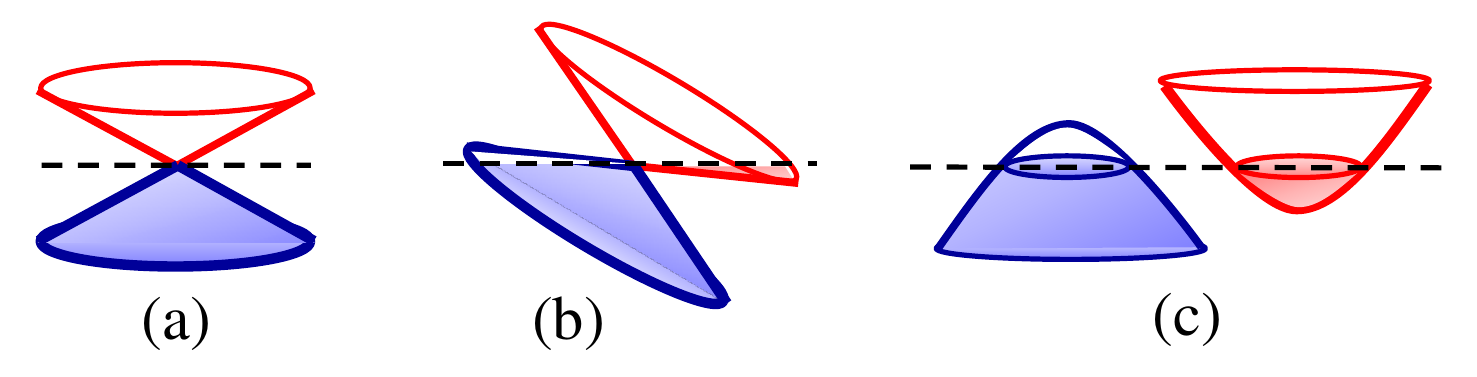}
\caption{(color online). Various types of semimetals. (a) Two conical bands that touch at a nodal point. The Fermi surface is a point. (b) Tilted cones (or type-II Weyl semimetal~\cite{Soluyanov2015}). The Fermi surfaces enclose an electron pocket and a hole pocket. (c) Overlapped bands that do not touch with each other. Dashed lines denote the location of the chemical potential. }
\label{fig:semimetal}
\end{figure}

In this paper, we show that, in sharp contrast to the chiral anomaly, Weyl nodes are in fact not vital for the existence of the CME. As shown in Fig.~\ref{fig:semimetal}, there are several ways to realize a semimetal. In Figs.~\ref{fig:semimetal}(a) and (b), when the nodes carry nonzero monopole charges, chiral anomaly and related transport phenomena are expected to emerge. In contrast, since there is no nodal point in Fig.~\ref{fig:semimetal}(c), chirality becomes meaningless and the monopole charge density (i.e., the divergence of Berry curvature) vanishes throughout the whole Brillouin zone. Therefore, there is no chiral anomaly in this nodeless case. However, we find that for clean samples the CME can exist even in this case with no Weyl node.

This conclusion may not be too surprising because, for both cases in Figs.~\ref{fig:semimetal}(b) and (c), energy bands are partially filled with finite Fermi surfaces, and the system is metallic in nature. Besides, in the semiclassical analysis,~\cite{Xiao_etal2010} there exists anomalous velocity induced by a Berry curvature: $\mathbf{v}_a = -\dot{\bf k}\times\mathbf{\Omega}_\mathbf{k}$, where $\hbar\dot{\bf k}$ is the force experienced by the electron. Under an external $\mathbf{B}$ field, the Lorentz force gives an anomalous velocity (and thus current): $\mathbf{v}_a \sim \frac{e}{\hbar}(\mathbf{v}\cdot\mathbf{\Omega}_\mathbf{k})\mathbf{B}$, where $\mathbf{v}$ is the group velocity of the Bloch electron. While this provides only a heuristic picture [the complete expression is presented in Eq.~\eqref{chmag_aniso}], it does show that the Berry curvature could drive a current along the direction of the $\mathbf{B}$ field, no matter whether the Berry curvature emanates from a monopole or not.

Our observation is illustrated by using a two-band lattice model. With suitable parameters, cases (a), (b), and (c) in Fig.~\ref{fig:semimetal} can all be realized. We find that, when two nodes merge in momentum space [see Fig.~\ref{fig:band}(a)] and the resulting node has zero total monopole charge, the CME would appear as long as there are electron and hole pockets at the Fermi level. That is, the energy separation between two nodes and nonzero monopole charge of the nodes are not necessary for the CME. This case is reminiscent of the type-II Weyl semimetal in Fig.~\ref{fig:semimetal}(b). Furthermore, for some parameters, the point degeneracy between energy bands can be lifted, so that there is no Weyl node. However, the CME still would appear when the energy bands are partially filled [see Fig.~\ref{fig:band}(b)]. Our results challenge the  wisdom based on the studies of linearized models, in which the chiral magnetic current arises from the presence of Weyl nodes and is proportional to the energy separation between Weyl nodes.

This paper is organized as follows: In Sec.~II, we introduce the two-band model under consideration. In Sec.~III, the dependence of the anomalous Hall conductivity and the CME coefficient on various parameters is studied numerically. The result of this work is summarized in Sec.~IV.

\section{Two-band lattice model for double-Weyl semimetal}
Here we consider the case of double-Weyl semimetals for illustration. Besides the usual linear Weyl nodes carrying monopole charges $Q=\pm 1$, there are nodal points with nonlinear dispersions and higher monopole charges.~\cite{Xu_etal2011,Fang_etal2012,Shivamoggi_Gilbert2013,Huang_a2015} Such nodes can be protected by crystallographic point group symmetries. One example of the so-called double-Weyl semimetal is HgCr$_2$Se$_4$, which contains nodes with $Q=\pm2$. The spectrum around each node disperses \emph{quadratically} in two directions. This material has recently been confirmed by a transport experiment to be a semimetal.~\cite{Guan_etal2015} It is predicted that, in a double-Weyl semimetal, the quantum anomalous Hall (QAH) conductivity, the coefficient of the chiral anomaly, and the number of Fermi arcs for surface states could all be doubled. The effect of electron interaction near such a quadratic node could also be very different from that of a linear node.~\cite{Moon_etal2013,Lai2015,Jian_Yao2015}

To study the CME of double-Weyl semimetals, we start with a two-band lattice Hamiltonian motivated by the compound HgCr$_2$Se$_4$,~\cite{Shivamoggi_Gilbert2013}
\begin{align}
H_0(\mathbf{k})=
&\left(\cos k_x - \cos k_y\right)\sigma^x + \sin k_x\sin k_y\sigma^y \nonumber\\
&+ \left(m - \cos k_x - \cos k_y - \cos k_z\right) \sigma^z \; ,
\label{dWeyl_Ham}
\end{align}
where $\sigma^\alpha$ ($\alpha=x$, $y$, $z$) are the three Pauli matrices
and $\mathbf{k}$ is the Bloch wavevector. This model breaks time-reversal symmetry, but has a combined $C_4$-rotation and $M_z$-mirror symmetry.
The bulk energy gap of Eq.~\eqref{dWeyl_Ham} closes and a pair of double-Weyl nodes emerges if $1<|m|<3$. When $1<m<3$, the nodes locate at $\mathbf{k}_0^\pm=(0,0,\pm\cos^{-1}(m-2))$; while for $-3<m<-1$, they locate at $\mathbf{k}_0^\pm=(\pi,\pi,\pi\pm\cos^{-1}(m+2))$ . Notice that the $C_4$ symmetry protects the double-Weyl points, while the $M_z$ symmetry requires that the two double-Weyl points have equal energy.

Since we are interested in studying the CME, a $M_z$-breaking term, $t_1 \sin k_z$, is added to split the energy of the two double-Weyl nodes. Besides, in order to split a double-Weyl node into two single-Weyl nodes, a $C_4$-breaking term, $a\,\sigma^x$, is added. Therefore, the lattice model under consideration becomes
\begin{align}
H(\mathbf{k})&= H_0(\mathbf{k}) + a\,\sigma^x + t_1 \sin k_z \nonumber\\
&=\mathbf{d}(\mathbf{k})\cdot{\bm\sigma}+ t_1 \sin k_z  \; ,
\label{whole_Ham}
\end{align}
where we have written $H$ in the standard $\mathbf{d}$-vector notation. The energy spectrum are invariant under the combined transformation of $a\to -a$, and $k_x \leftrightarrow k_y$. Therefore, it is sufficient to study the case with $a \geq 0$.

\begin{figure}
\includegraphics[width=3.0in]{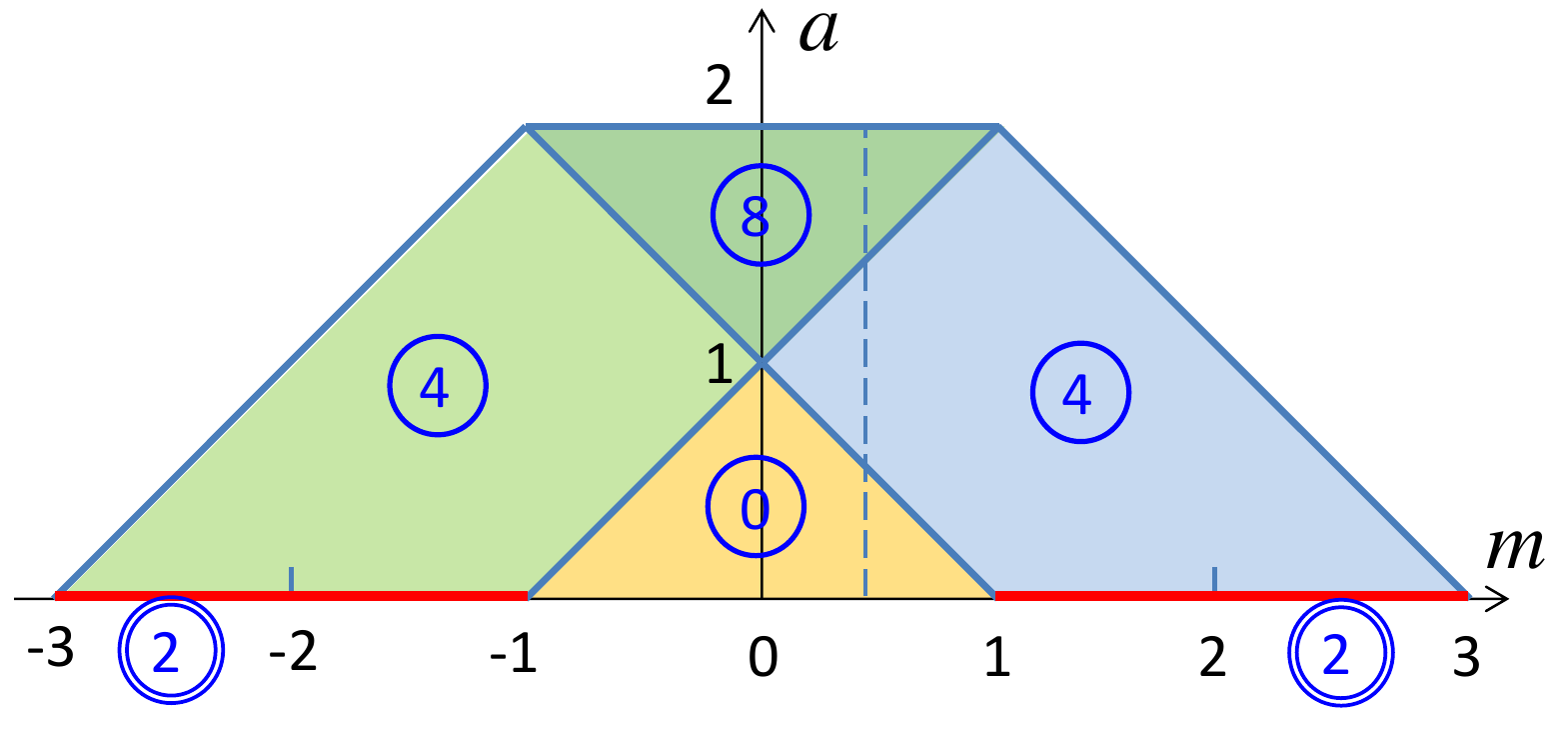}
\caption{(color online). Phase diagram of the Hamiltonian in Eq.~\eqref{whole_Ham} for $a\geq 0$. Here, a circled number (in blue) shows the number of Weyl nodes in a particular phase. Doubly circled numbers indicate that they are double-Weyl nodes (see Appendix A for more details).
This occurs when $a=0$ and $1<|m|<3$ (two red line segments). Outside the colored regions, the system is a trivial phase without node. The vertical dashed line indicates the path calculated in the right panels of Fig.~\ref{fig:data}.}
\label{fig:phase_diag}
\end{figure}

The phase diagram in the parameter space of $(m,a)$ is shown in Fig.~\ref{fig:phase_diag}. The $t_1$ term only tilts the band structure, but does not change the number of nodes in a phase. When $a=0$, as mentioned above, the system belongs to the semimetallic phase with a pair of double-Weyl nodes for $1<|m|<3$. If $1<m<3$, and $a\neq 0$, then each of the double-Weyl node would split to two nodes along the $k_x$-direction. The resulting phase has four linear Weyl nodes within the colored parallelogram in Fig.~\ref{fig:phase_diag}. Similarly, if $-3<m<-1$, then a positive $a$ would split a double-Weyl node to two nodes along the $k_y$-direction,
and we have another Weyl semimetal phase with four nodes. These two four-node phases would overlap to form a phase with eight nodes at larger $a$'s.

On the other hand, if $|m|<1$ and $a$ is small (the upright triangular region in Fig.~\ref{fig:phase_diag}), then there exists no node. When $t_1=0$ and the chemical potential $\mu=0$, the system becomes a three-dimensional QAH insulator with a quantized Hall conductivity $|\sigma_H|=2e^2/h$ (here the lattice constant is set to be unity).~\cite{Xu_etal2011,Moon_etal2013} The coefficient of two can be traced back to the fact that the monopole charge of a double-Weyl node is two times larger.

By choosing suitable parameters, both cases of tilted cones and overlapped bands in Figs.~\ref{fig:semimetal}(b) and (c) can be realized in the present model. For example, when $a=0$ and $m=1$, two (would-be) double-Weyl nodes merge at $\mathbf{k}_0^\pm=(0,0,\pi)$. Notice that the merged node carries no monopole charge. For a nonzero $t_1$ with a fixed chemical potential $\mu=0$, electron and hole pockets appear at the Fermi energy, as shown in Fig.~\ref{fig:band}(a). This resembles the type-II Weyl semimetal discussed in Ref.~\onlinecite{Soluyanov2015}. On the other hand, there is no node when the parameters $(m,a)$ lie within the upright triangular region in Fig.~\ref{fig:phase_diag}. Turning on a nonzero $t_1$, the energy bands are distorted so that the chemical potential $\mu=0$ intersects with both bands. A typical example is displayed in Fig.~\ref{fig:band}(b).

\section{Anomalous Hall effect and Chiral magnetic effect}
Within the linear-response theory, both the Hall conductivity and the CME coefficient can be calculated from the retarded current-current correlation function at finite frequency and wavevector, followed by an appropriate limiting process.~\cite{Chang_Yang2015}

For a generic two-band model, the Hall conductivity $\sigma^{ij}$ in the relation $J^i=\sum_{j\neq i}\sigma^{ij}E^j$ ($i,j=x$, $y$, $z$) is given as
\begin{equation}\label{AHE}
\sigma^{ij}=-\frac{e^2}{\hbar}\int \frac{d^3k}{(2\pi)^3} \; \sum_{t=\pm} \Omega^\ell_{\mathbf{k},t} \; f_t(\mathbf{k})  \; ,
\end{equation}
where $i,j,\ell$ are in the cyclic order of $x,y,z$. Here $f_t(\mathbf{k})$ is the Fermi-Dirac distribution function for band $t$($=\pm$) at some temperature $T$. The Berry curvatures $\Omega^i_{\mathbf{k},\pm}$  are given by the formula,~\cite{Bernevig}
\begin{eqnarray}\label{Berry}
\Omega^i_{\mathbf{k},\pm}= %
\pm \sum_{j,\ell}\epsilon^{ij\ell} \frac{1}{4d^3(\mathbf{k})} \; \mathbf{d}(\mathbf{k}) \cdot \left[ \frac{\partial \mathbf{d}(\mathbf{k})}{\partial k_j}\times \frac{\partial \mathbf{d}(\mathbf{k})}{\partial k_\ell}\right] \; .
\end{eqnarray}
Since the Weyl nodes in our model are connected by Dirac strings along the $z$-axis, the only non-vanishing components are $\sigma^{xy}=-\sigma^{yx}\equiv -\sigma_H$. The rest of the components are zero, which has been confirmed by numerical calculations.

\begin{figure}
\includegraphics[width=3.3in]{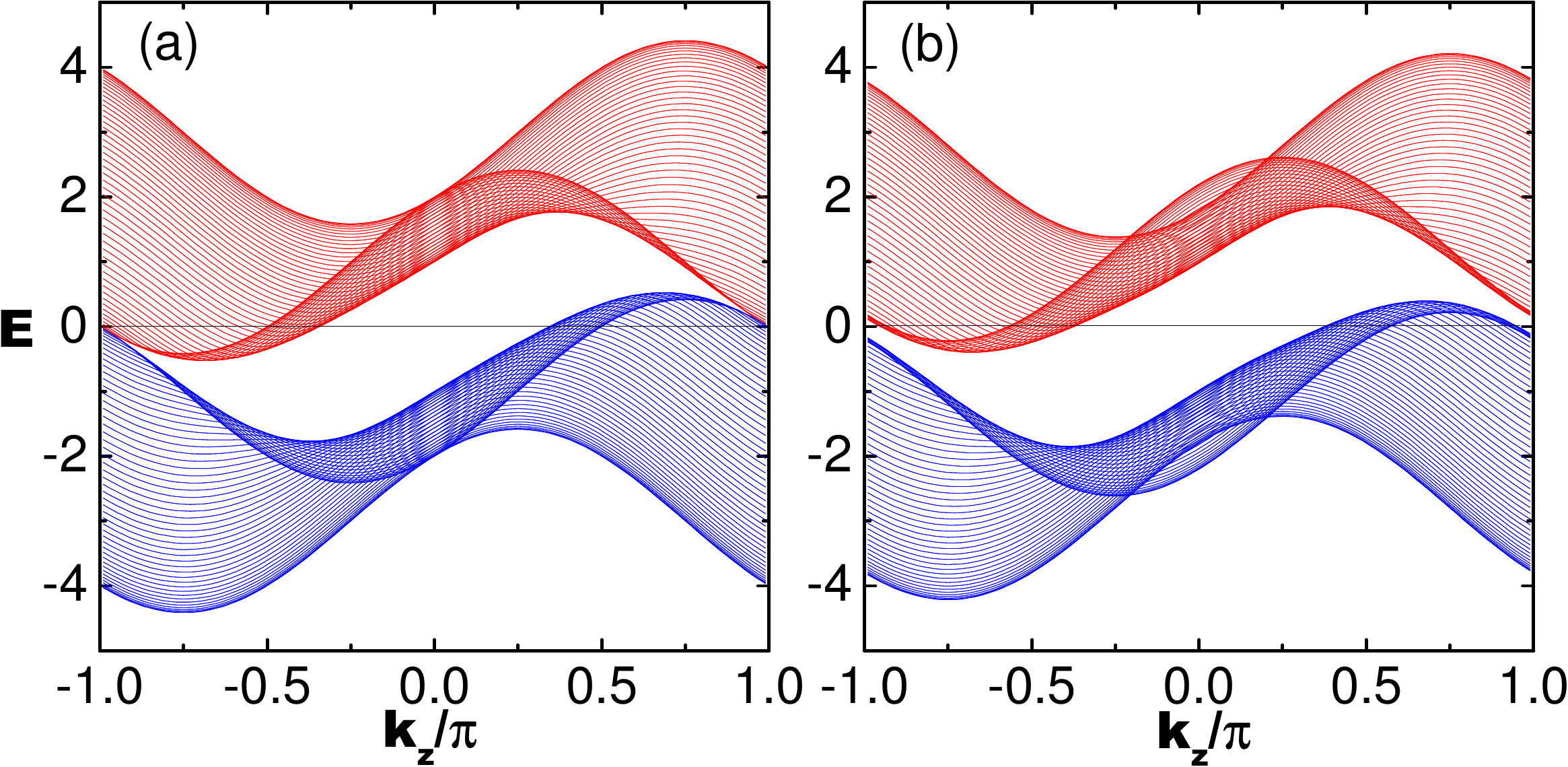}
\caption{(color online). Band energies as functions of $k_z$ ($k_x=k_y$). The parameters are (a) $m=1$ (a single merged node with \emph{zero} topological charge) and (b) $m=0.8$ (non-touching energy bands). Here $a=0$ and $t_1=1$.}
\label{fig:band}
\end{figure}

For a two-band model, following the steps described in Ref.~\onlinecite{Chang_Yang2015}, one can show that the CME coefficient for the chiral magnetic current $J_\mathrm{CME}^i=-\alpha^i B^i$ under an applied $\mathbf{B}$ field along direction $i$ is~\cite{note1}
\begin{eqnarray}\label{chmag_aniso}
\alpha^i &=&\frac{e^2}{\hbar}\int \frac{d^3k}{(2\pi)^3} \; \sum_{t=\pm} \left[\frac{\mathbf{v}_{\mathbf{k},+}+\mathbf{v}_{\mathbf{k},-}}{2} \cdot \mathbf{\Omega}_{\mathbf{k},t} \; f_t(\mathbf{k}) \right. \nonumber\\%
&& \left. %
- t\;d(\mathbf{k}) \; \frac{\mathbf{v}_{\mathbf{k},t} \cdot \mathbf{\Omega}_{\mathbf{k},t} - v^i_{\mathbf{k},t} \Omega^i_{\mathbf{k},t}}{2} \; \frac{\partial f_t}{\partial E_t}\right] \; .
\end{eqnarray}
Here $\mathbf{v}_{\mathbf{k},t}=(1/\hbar)\nabla_\mathbf{k}E_t(\mathbf{k})$ are the group velocities, and $E_t(\mathbf{k})$ are the band energies. Because the integrands of $\alpha^i$ along different directions differ only by a term containing $v^i_{\mathbf{k},t} \Omega^i_{\mathbf{k},t}$, the values of $\alpha^i$ do not differ much. They are often of the same order and show similar dependence on parameters. (This is true also for the model considered in Ref.~\onlinecite{Chang_Yang2015}.) For simplicity, only its averaged value, $\alpha=(\alpha_x+\alpha_y+\alpha_z)/3$, is presented below.

\begin{figure}
\includegraphics[width=3.3in]{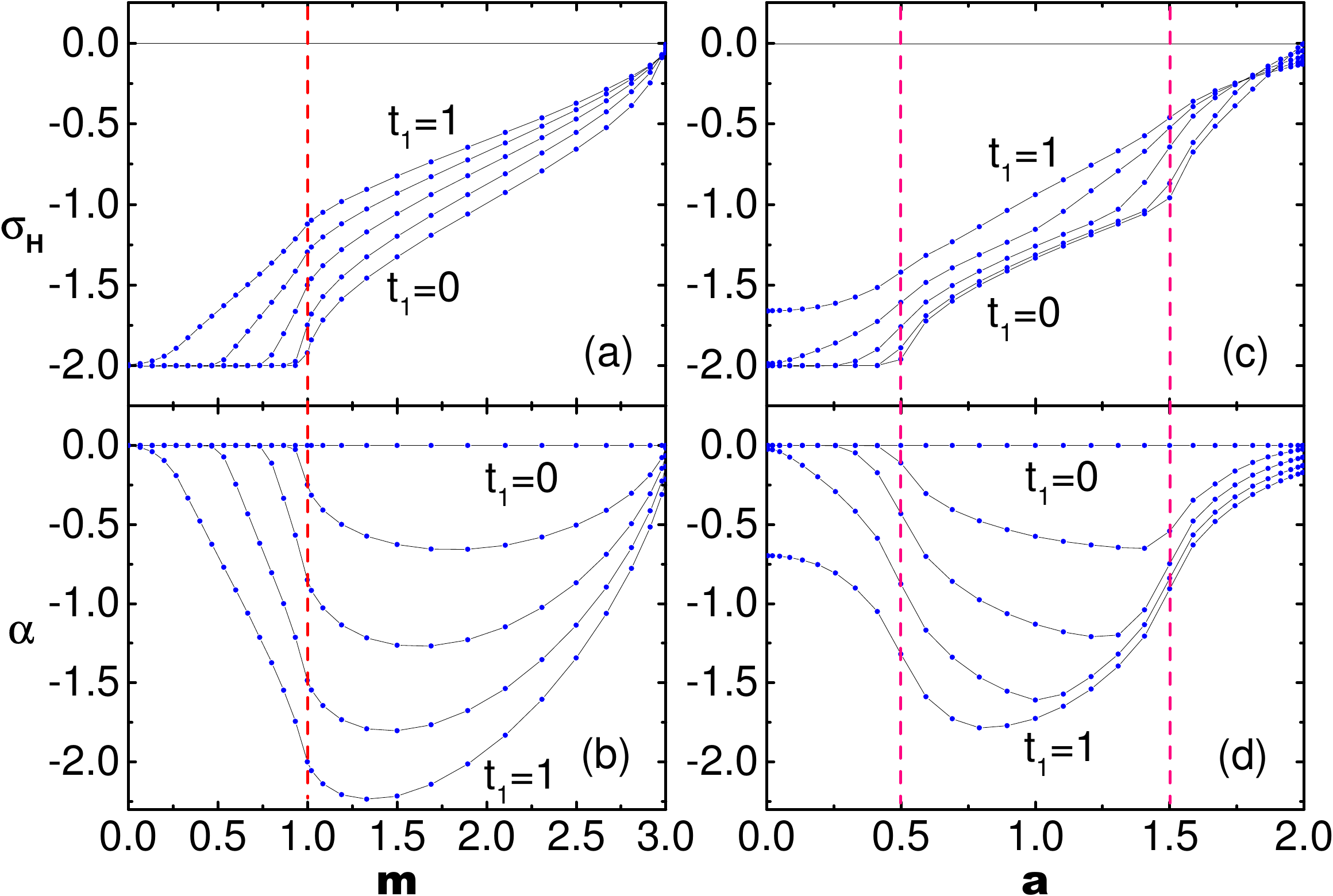}
\caption{(color online). Hall conductivity $\sigma_H$ (in units of $e^2/h$) and CME coefficient $\alpha$ (in units of $e^2/h^2$). Left panels: $a=0$ and $m\in [0,3]$ (along the positive $a$-axis in Fig.~\ref{fig:phase_diag}). Right panels: $m=0.5$ and $a\in [0,2]$ (along the vertical dashed line in Fig.~\ref{fig:phase_diag}). The parameter $t_1$ varies from 0 to 1 with step 0.25. Temperature $T=10^{-2}$ and the chemical potential $\mu=0$. Phase boundaries at $t_1=0$ are denoted by the vertical dotted lines. The calculations are done with $600^3$ lattice sites. }
\label{fig:data}
\end{figure}

The numerical results of  $\sigma_H$ and $\alpha$ for the two-band model in Eq.~\eqref{whole_Ham} are displayed in Fig.~\ref{fig:data}. We start with the $a=0$ case. When $t_1=0$, the system is nodeless and belongs to the QAH insulating phase with a quantized Hall conductivity $\sigma_H=-2e^2/h$ if $|m|<1$.~\cite{Xu_etal2011} As $m$ increases from 1 to 3, the two double-Weyl nodes approach each other, and thus the magnitude of $\sigma_H$ decreases to zero, when they annihilate each other at $m=3$. When $t_1\neq 0$, such that the valence band is not completely filled, $\sigma_H$ is no longer quantized even when $|m|<1$. These results are summarized in Fig.~\ref{fig:data}(a).~\cite{note2}

On the other hand, as seen from Fig.~\ref{fig:data}(b), the CME coefficient $\alpha$ vanishes as long as $t_1=0$. This is expected, because the system
is either in the insulating phase ($|m|<1$) or the nodes has no energy shift ($1<m<3$). Notably, when $t_1\neq 0$, $\alpha$ can be nonzero even if the system lies in the nodeless region of $|m|<1$. The finite value of $\alpha$ comes from the effect of the Berry curvatures of the electron and hole pockets [see Fig.~\ref{fig:band}(b)]. The value of $\alpha$ drops to zero at a smaller $m$ (a larger energy gap) as long as the chemical potential no longer intersects with the energy bands. Our results shows that the Weyl node is not required for the CME.

Now we consider the effect of the $C_4$-breaking term with $a\neq 0$. In Figs.~\ref{fig:data} (c) and (d), one follows the dashed line of $m=0.5$ in Fig.~\ref{fig:phase_diag} that starts in the phase with non-touching bands when $a=0$, enters the four-node phase when $a=0.5$, and reaches the eight-node phase when $a=1.5$. At $t_1=0$, there are kinks in the slope of the $\sigma_H$ curve at phase boundaries in Fig.~\ref{fig:data}(c).~\cite{note2} The curves become smooth when $t_1\neq 0$. Interestingly, as seen in Fig.~\ref{fig:data}(d), $\alpha$ behaves smoothly and non-monotonically when $t_1\neq 0$. Due to the same reason mentioned in the previous paragraph, $\alpha=0$ as long as $t_1=0$. On the two ends of $a=0$ and $a=2$, $\sigma_H$ and $\alpha$ can be non-zero at non-zero $t_1$, because the valence band is not completely filled. Most importantly, again the CME coefficient is obviously not zero in the nodeless region when $a<0.5$.

Lately, SrSi$_2$ is found to be a double-Weyl semimetal with pairs of nodes located at {\it different} energies.~\cite{Huang_a2015} Thus it could be a natural candidate for testing the CME.

\section{Conclusion and Discussion}
Even though the CME is originated from the theory of chiral fermions in high-energy physics, in condensed matters, it exists even in the absence of Weyl node and the chirality becomes ill-defined. Our analysis shows that the CME comes directly from the distribution of the Berry curvature in energy bands. Whether there are monopoles (sources/drains of the Berry flux) in the Brillouin zone is not crucial.

It worths emphasizing that the existence of the CME in a type-II Weyl semimetal or a nodeless phase is not restricted to the systems containing double-Weyl nodes. Same phenomena are found to exist (not shown here) also in a lattice model (an extended version of the one in Ref.~\onlinecite{Chang_Yang2015}) with only linear Weyl nodes. Because the CME coefficient is odd under space inversion, only the materials without inversion symmetry can exhibit the CME. Inspired by the present work, a candidate for realizing the nodeless CME could be a three-dimensional QAH insulator, after its energy bands being distorted by, e.g., inversion symmetry breaking, so that the valence (conduction) band is not completely filled (empty).

Recently, the chiral anomaly in Weyl semi-metal has been confirmed indirectly through the measurement of the negative longitudinal magnetoresistence (NLMR).~\cite{Huangx2015,Zhang2015} In the experimental setup, one applies electric and magnetic fields to a Weyl semi-metal. Because of the chiral anomaly, charges are pumped between Weyl nodes with opposite chiralities at a rate proportional to ${\bf E}\cdot{\bf B}$, which leads to a difference in chemical potentials between these Weyl nodes. Since the CME coefficient is proportional to such difference of chemical potentials, the electric conductivity is thus proportional to $B^2$, resulting in the NLMR. However, in this paper, we are studying the CME in an equilibrium state without an electric field. That is, the electric current is driven only by a magnetic field. Thus there is no connection between the CME here in an equilibrium state and the NLMR. In the case of nodeless CME, even if an additional electric field is applied, since there is no node, no chiral anomaly and associated charge pumping, so one does not expect to see the NLMR either.

Finally, we emphasize that the present linear-response analysis is done with a two-band model in a clean and infinite system. Real samples always have disorders and are finite in size. An important issue is how would these factors change the value of the CME coefficient. For example, the spin Hall conductivity for a clean Rashba system is first predicted to be non-zero, but later found to vanish with the inclusion of disorders.\cite{rashba} Similar qualitative alteration to the CME due to realistic factors cannot be completely ruled out. In any case, the conclusion in this paper should still hold in the conservative range of $\omega\tau\gg 1$, where $\omega$ is the driving frequency and $\tau$ is the relaxation time. Generalization to more realistic systems remains to be explored in future investigations.

\bigskip
\begin{acknowledgments}
M.C.C and M.F.Y. acknowledge the support from the Ministry of Science and
Technology of Taiwan under Grant NSC 102-2112-M-003-005-MY3 and NSC 102-2112-M-029-004-MY3, respectively.
\end{acknowledgments}

\appendix
\section{Calculating the monopole charge of a Weyl node}

The energy dispersion near a Weyl node determines its monopole charge.
As discussed in Ref.~\onlinecite{Felsager}, the monopole charge $Q$ (or the winding number of the mapping $f:\mathbf{k}\rightarrow\mathbf{d}$) of a given Weyl node can be calculated by a simple formula:
\begin{equation}
Q=\sum_{\ell=1}^N \mathrm{sgn} \left( \left|\frac{\partial d_i}{\partial k_j}\right|_{\mathbf{k}=\mathbf{k}^{(\ell)}} \right)  \; .
\end{equation}
Here one sums over the $\mathbf{k}$-points with $\mathbf{d}(\mathbf{k}^{(\ell)})=\mathbf{d}_0$, which is fixed, assuming that the Jacobian $|\partial d_i/\partial k_j|\neq 0$ at these ${\bf k}$-points.

For example, for $a=0$ and $m=2$ in our model, there are nodes at $\mathbf{k}_0^\pm=(0,0,\pm \pi/2)$. Expanding the momentum around the node, $\mathbf{k}=\mathbf{k}_0^\pm +\mathbf{q}$, one has
\begin{equation}
\mathbf{d}(\mathbf{q})\simeq \left(\frac{q_y^2}{2}-\frac{q_x^2}{2}, q_x q_y, \pm q_z\right),
\end{equation}
and the Jacobian is
\begin{equation}
\left|\frac{\partial d_i}{\partial q_j}\right|=\mp(q_x^2+q_y^2).
\end{equation}
Choosing $\mathbf{d}_0=(1/2,0,0)$, then there are two $\mathbf{q}$ points, $\mathbf{q}^{(1)}=(0,1,0)$ and $\mathbf{q}^{(2)}=(0,-1,0)$. They contribute to $Q=\mp2$ in total.



\begin{thebibliography}{99}
\bibitem{note1}
To be specific, this expression is derived under the \emph{uniform} limit, the wavevector $\mathbf{q}=0$ before the frequency $\omega\to 0$. In the opposite \emph{static} limit, $\omega=0$ before $\mathbf{q}\to 0$, there is a different expression for the response coefficient $\alpha$, which is found numerically to be zero. In convention, a transport coefficient is calculated under the uniform limit.
\bibitem{note2}
Due to thermal excitations at $T=10^{-2}$, our values of $\sigma_H$ at the phase boundary of the QAH insulating phase for $t_1=0$ deviate slightly away from $-2e^2/h$. The deviation becomes much smaller when $T=10^{-3}$. Such a finite-temperature effect has very little influence on the other data points.
\end{thebibliography}
\end{document}